\documentclass[10pt]{article}
\usepackage[utf8]{inputenc}
\usepackage{authblk}
\usepackage{biblatex}
\usepackage{xcolor}
\usepackage{float}
\usepackage{tikz}

\usepackage{graphicx}
\graphicspath{Images/}

\title{Rising Prevalence of Detected AI-Generated Text in Medical Literature: Longitudinal Analysis in Open Access Articles}

\author[1,2]{Nathan Wolfrath}
\author[1]{Simrin Patel}
\author[1]{Madelyn Flitcroft}
\author[1]{Anjishnu Banerjee}
\author[1]{Melek Somai}
\author[2]{Bradley H. Crotty}
\author[1,2]{Anai N Kothari}

\affil[1]{Bud and Sue Selig Hub for Surgical Data Science, Medical College of Wisconsin}
\affil[2]{Inception Health Labs, Medical College of Wisconsin }
\affil[3]{Department of Biostatistics, Medical College of Wisconsin}

\date{}

\begin{document}
\maketitle

\section{Introduction}
\noindent
Generative artificial intelligence (AI) tools have changed the way text can be produced across domains, including academic writing. The rapid adoption of both commercial and open-weight large language models (LLMs) has raised new questions about the authenticity and provenance of scholarly content as well as the proper role for AI in the publication process. The widespread availability of these tools has prompted many academic publishers to release mandatory disclosures and other related policies to guide use of AI in writing. However, very little is known about the extent to which AI-generated content actually appears in peer-reviewed medical literature. 

In parallel with the rise of large language models, various tools and methods have been developed by both academics and commercial entities alike for detecting AI generated text. Approaches have ranged widely in complexity from purely statistical analysis to deep learning and transformer-based methods\cite{wu-etal-2025-survey}. Due to the rapidly changing landscape of generative AI and the rate of new LLM releases, the accuracy of these tools remains a topic of ongoing investigation. However, increasing efforts have been made to rigorously characterize the performance of detectors in a standardized way\cite{dugan2024raid,wu2024detectrl}. 

In this brief study, we aim to assess the prevalence and trends of AI-generated writing in high-quality open access medical literature using commercial detection software. We examine both absolute rates and trends over the analyzed time period from 2022 to early 2025. Furthermore, we compare AI use disclosed by authors to the rate of detected use.

\section{Methods}
\noindent
The main body plain text of all Original Investigation, Research Letter, and Invited Commentary publications in JAMA Network Open from January 2022 through March 2025 were obtained through the journal website via web scrape. Permission was granted from the journal for this use case. JAMA Network Open was selected for the number of articles, the varied domains with standardized classification, the availability of open-access complete manuscripts, and the high impact factor \cite{jamanetwork}.

Web scraping was performed using Python. After retrieval, each web page was parsed with the Beautiful Soup to extract manuscript body text as well as publication month, publication format, and domain from article metadata. In-text reference numbers were programatically removed. Regular expression-based matching was used in the acknowledgment section (where AI disclosures appear on the JAMA Network website) of each article to determine whether a language model was disclosed to aid in the writing process. Search terms included ChatGPT, GPT, Claude, Gemini, Llama, Mistral, Mixtral, Sonnet, ChatGLM, and Grok with word-boundaries. Matches were manually reviewed for validity. Acknowledgments which noted LLM use in writing analytic code, figure generation, or other applications outside of drafting or editing the manuscript text were not included. 

Originality.AI \cite{originality-ai}, a commercial AI detector, was used for detection of AI-generated text. This tool has been previously assessed in academic works and consistently performs well relative to other detectors\cite{han2025teachers,howard2024characterizing,flitcroft2024performance}. Each article was analyzed to compute the probability that the manuscript included greater than 10\% AI-generated text (the threshold defined by the model manufacturer). Probabilities were transformed into a binary label at a threshold of 0.5, as recommended by Originality.AI. The resulting data was aggregated by publication month for time series analysis and by publication type and domain for further comparisons. Chi-squared testing was used to determine whether distributions differed significantly across publication type and domain. Mann-Kendall testing was used for monotonic trend detection over the analyzed time period.

\section{Results}
\noindent
7251 articles were retrieved and analyzed, of which 195 (2.7\%) were classified as containing AI-generated text over the entire analyzed time period. The proportion of published articles including AI-generated text each month generally increased over time though with some variation. This included an increase from 0.0\% in the first month analyzed (January 2022) to 11.3\% in the final month (March 2025). There was a significant increasing trend over this period (P$<$0.001).  

\begin{figure}[H]
    \centering
    \includegraphics[width=\textwidth]{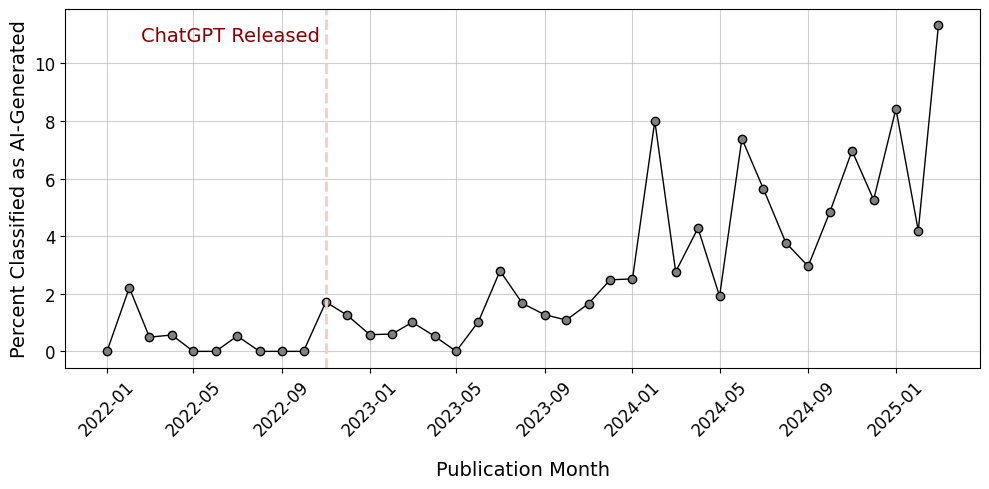}
    \caption{Percent of manuscripts classified as AI-influenced monthly since January 2022. ChatGPT was released by OpenAI on November 30, 2022.}
    \label{fig:fig1}
\end{figure}

There was significant variation in detected LLM text across publication types (P$<$0.001), with Invited Commentary having the highest proportion of articles classified as including AI-generated text (6.7\%), followed by Original Investigations (2.2\%) and Research Letters (1.4\%). Variation across domain was also significant (P=0.04), with Global Health (8.6\%), Genetics and Genomics (5.7\%), and Medical Education (5.6\%) having the highest percentage of publications classified as containing AI-generated text and several fields having all articles classified as exclusively human-written (Figure 2).

\begin{figure}[H]
    \centering
    \makebox[\textwidth][c]{%
        \includegraphics[width=0.8\paperwidth]{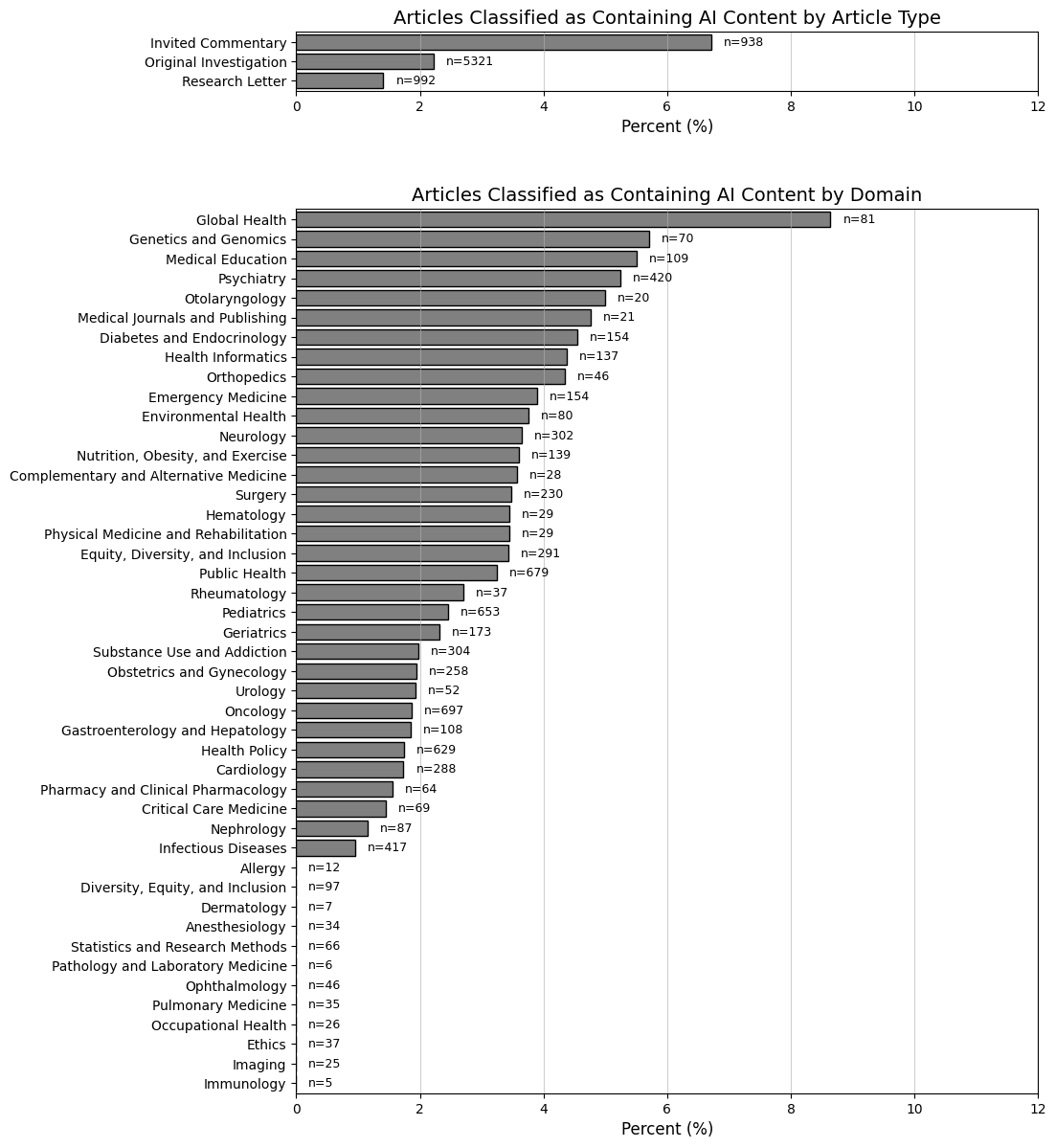} 
    }
    \caption{Percent of manuscripts classified as AI-influenced since January 2022 stratified by publication type and domain. N represents total articles analyzed for each domain.}
    \label{fig:fig1}
\end{figure}

Of all articles, fifteen articles (0.2\%) were found to disclose utilizing an LLM in the writing process, six (40.0\%) of which were classified as over 10\% AI-generated. Of all 195 articles with text detected as AI generate, this represents 3.1\% which disclosed use of AI in the writing process.   

\section{Discussion}
This analysis demonstrates an upward trend in content detected as AI-generated following the release of ChatGPT in November 2022, with the highest prevalence in Invited Commentaries. Across domains, Global Health had the most content considered AI-generated. It is hypothesized writers who are not primarily English-speaking requiring LLMs for writing assistance, though writing in a second language may affect detector performance \cite{liang2023gpt}. A separate analysis based on author language was not attempted as the author’s native language cannot be reliably determined without inaccurate proxies such such as institution location. 

As of March 2025, 11.2\% of publications were classified as having AI-generated text. Despite JAMA Network Open’s policy for disclosing AI use, only a small number of published manuscripts included this acknowledgment. As generative and detection models continue to evolve, continued characterization of performance and reliability of both will be necessary to inform and enforce policies guiding the use of AI in academic writing. 

Importantly, AI detection tools have shortcomings, so results must be interpreted carefully. For example, articles flagged as containing AI-generated content prior to the popularization of generative AI tools occurred at a monthly in 0-2.2\% of analyzed manuscripts. These most likely represent false positives, though it is possible tools preexisting current LLMs such as article spinners, grammar correction tools, or translation software could influence these classifications. 

Of papers which note LLM use, 40\% were classified as positive by the detector. The extent of LLM use is unknown and may have been insufficient to meet the 10\% threshold, or alternatively represent false negatives. Although usage was disclosed these works, the true proportion of LLM generated content is not known. Due to this and the small sample size of disclosures, further calibration of the detection tool based on this subgroup was not attempted. 

Finally, detectors may have significant variance in performance based on factors including different language models, model parameters, context, specific use (e.g. editing or rewording versus de novo generation), and detection evasion strategies \cite{fraser2025detecting,sadasivan2023can}. Therefore, these results cannot be considered a definitive description of what proportion of publications or specific articles have utilized AI, but rather an estimate using the best available technology at this time.  

\section{Acknowledgments}
GPT-4o was used in April-May 2025 to assist with writing code for analysis and figure generation. Output was validated by human expert review and the authors take responsibility for its accuracy. All data (text manuscripts) used in this study are open access, and we thank \textit{JAMA Network Open} for granting permission for this use case. The authors recieved no funding for this work and have no financial relationship or other affiliation with OpenAI or Originality.AI.

\nocite{*}  

\printbibliography

\end{document}